\documentclass[aps,prd,floatfix,notitlepage,superscriptaddress,preprintnumbers,nofootinbib,10pt]{revtex4-1}
\usepackage{graphicx}
\usepackage{amssymb}
\usepackage{amsmath,amssymb}
\usepackage{color}
\usepackage{setspace}
\usepackage{fancyhdr}
\usepackage{multirow,natbib}
\definecolor{darkblue}{rgb}{0,0,0.5}
\usepackage[colorlinks,linkcolor=blue,citecolor=blue,urlcolor=blue]{hyperref}

\allowdisplaybreaks

\newcommand\beq{\begin{equation}}
\newcommand\eeq{\end{equation}}

\newcommand\nn{\nonumber}

\newcommand\hc{\text{h.c.}}

\def\Y{\hat{Y}}

\def\sol{\rm sol}
\def\atm{\rm atm}
\def\tbeta{\tan\beta}

\usepackage{titlesec}
\begin{document}



\title{Zee model with quasidegenerate neutrino masses and where to find it}

\author{R. Primulando}
\email{rprimulando@unpar.ac.id}
\affiliation{Center for Theoretical Physics, Department of Physics, Parahyangan Catholic University, Jl. Ciumbuleuit 94, Bandung 40141, Indonesia}

\author{J. Julio} 
\email{julio@brin.go.id}
\affiliation{National Research and Innovation Agency, Kompleks Puspiptek Serpong, South Tangerang 15314, Indonesia}

\author{P. Uttayarat}
\email{patipan@g.swu.ac.th}
\affiliation{Department of Physics, Srinakharinwirot University, 114 Sukhumvit 23 Rd., Wattana, Bangkok
10110, Thailand}

\begin{abstract}
We present a Zee model with a family dependent $Z_2$ symmetry for radiative neutrino masses. Our motivation is to get a model that correctly describes neutrino oscillation phenomena, while at the same time offers definite predictions. The imposed $Z_2$ symmetry greatly reduces the number of free parameters in the model. These parameters are then determined from the neutrino data, from which one can study its outcomes. Our setup only admits quasidegenerate neutrino masses with the sum of neutrino masses between 100 and 130 meV, the effective Majorana mass between 20 and 40 meV, and the effective electron neutrino mass between 48 and 53 meV. The ratio of the vacuum expectation values of the Higgs doublets, $\tbeta$, is found to be $\tbeta\lesssim 0.5$ and $\tbeta\gtrsim 10$. The former is ruled out by lepton flavor violation (LFV) processes, such as $\mu\to e\gamma$ and $\mu\to e$ conversion, which are determined up to two loops.  For the latter, these LFV processes are within reach of the next generation of experiments. Moreover, for $\tbeta\gtrsim10$, the couplings of heavy neutral scalars to dimuon are significant. If they are  sufficiently light, i.e., $\lesssim200$ GeV, collider search for their decays into muon pair provides a stronger constraint on most parts of the model parameter space than the LFV ones. 
\end{abstract}

\maketitle

\flushbottom
\section{Introduction}
The mounting evidence of neutrino oscillations has led to the fact that neutrinos must be massive and mix. This indicates that the Standard Model (SM) must be considered as an effective theory; it cannot explain how neutrinos gain masses, let alone why their masses are so tiny compared to those of other SM fermions. There is a variety of models in literature attempting to explain this phenomenon. The simplest example  is the (type-I) seesaw mechanism, in which neutrinos get their masses at tree level via an exchange with electroweak singlet right-handed neutrinos \cite{Minkowski:1977sc,GellMann:1980vs,Yanagida:1979as,Mohapatra:1979ia,Glashow:1979nm}. The correct neutrino masses of order $0.1$ eV will require the right-handed neutrino masses to be very large, i.e., of the order $10^{14}$ GeV. With such a large scale, it is then difficult to test this mechanism at laboratories.   

There is also another alternative of neutrino mass generation, namely the radiative mechanism. Here neutrino masses are induced at loop level, see \cite{Zee:1980ai,Zee:1985id,Babu:1988ki} for pioneers of this class of models. Within this mechanism, due to loop and, possibly, chirality suppression factors, the new physics scale can be much lower; in some cases it can be even at TeV scale, placing it within reach of the Large Hadron Collider (LHC), see for example Refs. \cite{Zee:1985id,Babu:1988ki,AristizabalSierra:2006gb,Nebot:2007bc,Babu:2015ajp,Babu:2010vp}. One of the well-studied models within this class is the Zee model~\cite{Zee:1980ai}. Considered as an extension of the two-Higgs-doublet model by a singly-charged scalar, in its general form, this model induces tree-level flavor-changing neutral currents (FCNCs) in both lepton and quark sectors. Usually, such dangerous processes can be avoided by imposing the natural flavor conservation~\cite{Glashow:1976nt,Paschos:1976ay}, by which only one Higgs doublet can couple to right-handed fermions of the same charge. However, if one applies this to neutrino mass generation~\cite{Wolfenstein:1980sy}, it will result in vanishing of all diagonal elements of neutrino mass matrix, leading to its exclusion by solar neutrino and KamLAND data. It should be noted, however, that the general Zee model is still compatible with oscillation data, but owing to its large number of parameters, it is hard to make a prediction out of it. For earlier phenomenological studies of the general Zee model, see Ref.~\cite{Petcov:1982en,Smirnov:1996bv,Jarlskog:1998uf,Frampton:1999yn,Koide:2000jm,Ghosal:2001ep,Babu:2013pma,Babu:2019vff,Babu:2019mfe,Nomura:2019dhw}.

In the past, an attempt to make a viable and predictive Zee model was done by introducing a family-dependent $Z_4$ symmetry~\cite{Babu:2013pma}. Within this scenario, tree-level FCNCs in the lepton sector were allowed, but their rates were found to be below experimental limits  even for very light scalar mediators. Thanks to the reduced number of parameters, the model was very predictive: it admitted only the inverted ordering of neutrino mass with no CP violation in the lepton sector. This result was in an excellent agreement with neutrino data for some time, but now it has been disfavored by the T2K result~\cite{T2K:2019bcf}, which has reported the evidence of nonconserving leptonic CP violation at more than $3\sigma$ for the IO case. Although other experiment such as NOvA may not severely exclude the CP conserving region \cite{NOvA:2021nfi,NOvA:2019cyt}, it is nevertheless useful to seek a model that can naturally allow for CP violation.

In this paper, we modify the Zee model by imposing a family-dependent $Z_2$ symmetry. This allows a nontrivial mixing in the lepton sector, hence tree-level FCNCs. The group itself is smaller than $Z_4$, so one may expect that the current scenario contains more parameters than those using $Z_4$. In fact, here we have an additional complex parameter compared to that of Ref.~\cite{Babu:2013pma}.  It makes the neutrino mass matrix complex, which in turn allows for nonvanishing CP violation. In total, this model contains 6 real parameters describing the neutrino mass and mixing. All can be fitted into neutrino oscillation data well, with a prediction that the present model accommodates only the quasidegenerate neutrino masses. As a consequence,  the result of this model is very sensitive to future experiments aiming on measuring neutrino mass. The sum of neutrino masses  and  the effective mass for neutrinoless double beta decay, for example, are estimated to be within the range of 100 to 130 meV and 20 to 40 meV, respectively. Both ranges are accessible to experiments such as Simon Observatory~\cite{SimonsObservatory:2019qwx} and LEGEND~\cite{LEGEND:2017cdu}.

Our model is also constrained by the collider and lepton flavor violation (LFV) bounds, which are dependent of $\tan\beta$, i.e.,  the ratio of the two vacuum expectation values of the Higgs doublets. Its value is determined solely from the neutrino data fit, in which we find that $\tan\beta$ is clustered into two regions, namely,  $\tbeta\lesssim 0.5$ and $\tbeta\gtrsim 10$, independent of the detail of the scalar phenomenology. The LFV processes, such as $\mu \to e\gamma$ and $\mu\to e$ conversion in nuclei, evaluated up to the two--loop Barr--Zee diagrams~\cite{Barr:1990vd}, further rule out the $\tan\beta\lesssim 0.5$ regime for wide range of scalar masses (i.e., $\lesssim 10$ TeV). This is partly because the corresponding $e\mu$ coupling is significantly enhanced in this case. Meanwhile, the Higgs couplings to top quarks, relevant for such two--loop diagrams (see~\cite{Davidson:2010xv,Harnik:2012pb}), remain sizable. This is in contrast to the case of $\tan\beta\gtrsim 10$, where the corresponding LFV rates are found to be well below the current bounds, and thus can only be probed by future generation experiments, such as Mu2e~\cite{Bernstein:2013hba} or MEGII~\cite{MEGII:2018kmf}. 
But what we find interesting is the importance of collider search. Specifically we find that in one of the two possible ways to couple quarks to the Higgses, the decay of a heavy Higgs into a dimuon final state gives a stronger constraint on the model than the present LFV transitions, provided that the heavy Higgs is 200 GeV or lighter.
Using the CMS partial data set of 35.9 $\rm fb^{-1}$  on this particular $\mu^+\mu^-$ channel~\cite{CMS:2019mij}, we can rule out a great portion of the model parameter space. The CMS full data set, once available, will certainly be able to falsify this scenario.

The rest of the paper is organized as follows. In Sec.~\ref{sec:model}, we shall discuss the detail of the model. In Sec.~\ref{sec:neutrino}, we discuss how this model describes the neutrino mass and mixing. This is followed by the discussion of relevant lepton flavor violation processes in Sec.~\ref{sec:lfv} and collider phenomenology in Sec.~\ref{sec:heavyhiggs}. We conclude and discuss our finding in Sec.~\ref{sec:conclusion}.

\section{The model}
\label{sec:model}

The model presented here is an extension of the SM. The scalar sector is expanded to contain two Higgs doublets $\Phi_a$ with $a=1,2$ and a singly-charged scalar singlet $\eta^+$. A discrete $Z_2$ symmetry that is flavor dependent is imposed on the lepton and the scalar sectors as follows
\begin{equation}
	L_i\equiv\begin{pmatrix} \nu_i \\ \ell_i \end{pmatrix}:(-1,1,1),\quad e_{R\,i}: 1,\quad \Phi_1: 1,\quad \Phi_2:-1,\quad \eta^+:-1,
	\label{eq:z2}
\end{equation}
where $L$ and $e_R$ denote the SM lepton doublet and singlet, respectively, and $i$ is a lepton-family index $i=e,\mu,\tau$.  Family-dependent $Z_2$ charge assignment in the lepton sector will lead to LFV processes. We will consider LFV constraints in Sec.~\ref{sec:lfv}. 

With the charge assignment given in Eq.~\eqref{eq:z2}, the Yukawa interactions in lepton sector can be written as
\begin{equation}
	{\cal L}_{\rm Yuk}^{\rm lepton} = \sum_{i=e,\mu,\tau}\sum_{\alpha=\mu,\tau}Y_{\alpha i}\bar L_\alpha e_{R i}\Phi_1 + \sum_{i=e,\mu,\tau}Y_{i}\bar L_e e_{Ri}\Phi_2 + \sum_{\alpha=\mu,\tau}f_{e\alpha}L_e^TC(i\sigma^2)L_\alpha\eta^+  +\hc\,,
	\label{eq:yukawa}
\end{equation}
 where $C$ is the charge conjugation matrix and the $SU(2)_L$ indices are contracted by the antisymmetric tensor $i\sigma^2$.
Due to the Fermi statistics, the coupling matrix $f$ is antisymmetric in flavor space. 
In addition to the aforementioned Yukawa interactions, there also exists a gauge-invariant term that is part of the scalar potential
\begin{align}
V \supset \mu \Phi_1^T(i\sigma_2)\Phi_2\eta^- + \hc
\label{mu-trilinear}
\end{align}
The non-vanishing $f$ and $\mu$ break lepton number by  two units, leading to the generation of Majorana-type neutrino masses. 

After the electroweak symmetry breaking, each Higgs doublet acquires vacuum expectation value (vev), denoted as $\left<\Phi_a\right>\equiv v_a/\sqrt{2}$. It is also customary to rotate the two Higgs doublet into the so-called Higgs basis
\begin{align}
&H_1 =\cos\beta \Phi_1 + \sin\beta \Phi_2 \nonumber \\
&H_2 = \sin\beta \Phi_1 - \cos\beta\Phi_2,
\label{eq:rotation-Higgs}
\end{align}
with $\tan\beta \equiv v_2/v_1$, such that only one Higgs doublet has vev $v=\sqrt{v_1^2+v_2^2}=246~{\rm GeV}$. In a more explicit form we have
\begin{equation}
	H_1 = \begin{pmatrix}G^+\\[1em] \dfrac{v+h_1+iG}{\sqrt{2}} \end{pmatrix},\quad
	H_2 = \begin{pmatrix}H^+ \\[1em] \dfrac{h_2+iA}{\sqrt{2}}\end{pmatrix}.
	\label{eq:doublet}
\end{equation}
Here $G^+$ and $G$ are the would-be Goldstone bosons eaten by the $W^+$ and $Z$, while $H^+$ and $A$  are the massive charged scalar and  CP-odd scalar, respectively. The two CP-even scalars, denoted by $h_1$ and $h_2$, in principle, can mix with the CP-odd one. However, throughout this paper, we shall assume that CP be a good symmetry of the scalar sector, so there is no mixing between CP-even and CP-odd scalar fields. Furthermore, based on current measurements of the 125 GeV Higgs couplings at the LHC~\cite{Khachatryan:2016vau,CMS:2018uag,ATLAS:2019nkf}, which are in good agreements with the SM expectations, it is instructive to adopt the decoupling limit, in which $h_1$ couples to fermions and gauge bosons as in the SM. Throughout this paper, we shall identify $h_1$ with $h$, the 125 GeV Higgs boson, and $h_2$ with $H$, the heavy Higgs. It is straightforward to generalize our result to a case where $h_1$ can mix with $h_2$.

\begin{figure}[t]
\begin{center}
\includegraphics[width=0.45\textwidth]{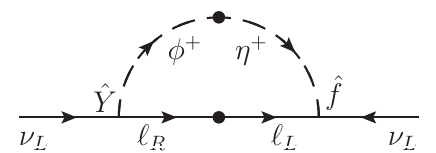}
\caption{The one-loop diagram generating neutrino masses.}
\label{fig:numass}
\end{center}
\end{figure}

It is worth noting that, due to the trilinear coupling of Eq.~\eqref{mu-trilinear}, there is also a mixing between $H^+$ and $\eta^+$ characterized by
\begin{equation}
	\begin{pmatrix}H^+\\\eta^+\end{pmatrix} = \begin{pmatrix}\cos\gamma &-\sin\gamma\\\sin\gamma &\cos\gamma\end{pmatrix}\begin{pmatrix}H_1^+\\H_2^+\end{pmatrix},
\end{equation}
where $H_{1,2}^+$ are mass eigenstates and $\sin2\gamma=\sqrt{2}\,\mu v/(m_{H_1^+}^2-m_{H_2^+}^2)$.
However, such mixing is constrained by neutrino mass to be rather small, see Fig.~\ref{fig:numass}. Therefore, in practice, we can treat both charged scalars as nearly physical fields. 

In the Higgs basis, the Yukawa interactions involving the Higgs doublets read 
\begin{align}
{\cal L}_{\rm Yuk}^{\rm lepton} \supset \sum_{i=e,\mu,\tau}\sum_{\alpha=\mu,\tau}Y_{\alpha i}\bar L_\alpha e_{R i}\left[ \cos\beta H_1 + \sin\beta H_2 \right] + \sum_{i=e,\mu,\tau}Y_{i}\bar L_e e_{Ri} \left[ \sin\beta H_1 - \cos\beta H_2 \right],
\label{eq:leptonyuk1}
\end{align}
The flavor structures of these two Yukawa matrices, $Y_i$ and $Y_{\alpha i}$,  bear similarity to the ones presented in Ref.~\cite{Babu:2013pma}. The latter can be considered as a $3\times 3$ matrix with the vanishing first row. We can bring such a matrix into a diagonal form, with $Y_{\mu\mu}$ and $Y_{\tau\tau}$ being the only nonzero entries, by performing a unitary field redefinition on the $L_{\mu,\tau}$ and $e_{R i}$ fields. Note that such field redefinitions affect neither the form of the antisymmetric coupling matrix $f$ nor the form of gauge charged currents. The former keeps its antisymmetric form
\begin{align}
f = f_{e\mu} \begin{pmatrix} 0 & 1 & re^{i\theta} \\
-1 & 0 & 0 \\
-r e^{i\theta} & 0& 0 
\label{f-explicit}
\end{pmatrix},
\end{align}
with $re^{i\theta}\equiv f_{e\tau}/f_{e\mu}$ because the field redefinitions are performed on the $\mu$-$\tau$ sector.

Using the basis where $Y_{\alpha i}$ is diagonal, the lepton mass matrix can be expressed as
\begin{equation}
	M_\ell = \dfrac{v}{\sqrt{2}}\begin{pmatrix} Y_{e} \sin\beta &  Y_{\mu} \sin\beta &  Y_{\tau}\sin\beta \\ 0 & Y_{\mu\mu}\cos\beta  &0 \\0 &0 & Y_{\tau\tau} \cos\beta \end{pmatrix}.
\end{equation}
Similar to the case of Ref.~\cite{Babu:2013pma}, all complex phases in the above leptonic mass matrix can be removed, leaving us with a real and positive matrix. We, then, diagonalize $M_\ell$ by a bi-orthogonal transformation
\begin{equation}
	O_L^TM_\ell O_R^{\phantom{T}} = \text{diag}(m_e,m_\mu,m_\tau),
	\label{mleptond}
\end{equation}
where $O_L$ and $O_R$ are orthogonal matrices used to rotate the $\ell_i$ and $e_{Ri}$ fields into their mass eigenstates. They can be written as
\begin{align}
O_L = \begin{pmatrix}
c_{\theta_3} & -s_{\theta_3} & 0 \\ s_{\theta_3} & c_{\theta_3} & 0 \\ 0 & 0 & 1
\end{pmatrix}
\begin{pmatrix}
c_{\theta_2} & 0 & -s_{\theta_2} \\ 0 & 1 & 0 \\ s_{\theta_2} & 0 & c_{\theta_2}
\end{pmatrix}
\begin{pmatrix}
1 & 0 & 0 \\ 0 & c_{\theta_1} & -s_{\theta_1} \\ 0 & s_{\theta_1} & c_{\theta_1}
\end{pmatrix},
\label{eq:ol}
\end{align}
where $c_{\theta_i} (s_{\theta_i})$ stands for $\cos\theta_i (\sin\theta_i)$. For $O_R$, one just  needs to make a replacement $\theta_i \to \alpha_i$. Now, out of 5 real parameters in  $M_\ell$, 3 of them will become charged lepton masses, indicating that only two mixing angles turn out to be physical. There is no sacred recipe of how to choose these physical angles. One can, for example, use the four zero entries of $M_\ell$ in Eq.~\eqref{mleptond} to eliminate four angles in favor of the two physical ones. These two angles, say, $\theta_1$ and $\alpha_3$, will later be determined through neutrino oscillation data. We give the complete expression in Appendix~\ref{app:diag}. Note that the procedure outlined above, in conjunction with taking the decoupling limit, will make leptonic interactions with $h$ flavor diagonal. Their strengths are proportional to lepton masses, as in the SM.

Having diagonalized the charged lepton mass matrix, we now turn to flavor-changing Yukawa interactions, which in the present case are given by
\begin{align}
\mathcal{L}_{\rm Yuk}^{\rm lepton} \supset f_{ij}L_i^TC(i\sigma^2)L_j\eta^+ + \sqrt{2}Y_{ij} \bar L_i e_{Rj} H_2 + \hc.
\label{eq:leptonyukawa}
\end{align}
The explicit form of  $f$ can be seen in Eq.~\eqref{f-explicit}, while the matrix coupling $Y$ is defined as
\begin{align}
Y = \dfrac{1}{v}\begin{pmatrix}
-\cot\beta & 0 & 0 \\
0 & \tan\beta & 0 \\
0 & 0 & \tan\beta 
\end{pmatrix} M_\ell.
\label{Y-exp}
\end{align}
Following the rotation of the leptonic fields into their mass eigenbasis,  the two matrices $f$ and $Y$ get transformed into $\hat f = O_L^TfO_L^{\phantom{T}}$ and $\hat Y = O_L^T Y O_R^{\phantom{T}}$, respectively. Such rotation induces the (2,3) entries in the antisymmetric coupling $\hat f$, which were zero in the original basis. As for $\hat Y$, Eq.~\eqref{Y-exp} implies that  it has a hierarchical structure, namely $\hat Y_{ij}/\hat Y_{ji} = m_{\ell_j}/m_{\ell_i}$. Now with all this in hand, we can write down  the Yukawa interactions in lepton sector as follows  
\begin{equation}
{\cal L}_{\rm Yuk}^{\rm lepton} = 2\sum_{i<j}\hat{f}_{ij} \left(\nu^T_i C \ell_j - \nu_j^T C \ell_i \right) \eta^+ + \dfrac{m_{\ell_i}}{v}\delta_{ij}\bar\ell_i e_{Rj} h
+ \sqrt{2}\hat Y_{ij}\bar\nu_i e_{Rj}H^+ + i\hat Y_{ij}\bar\ell_i e_{Rj} A + \hat{Y}_{ij} \bar\ell_i e_{Rj}H     +\hc\,.
	\label{eq:leptonyuk}
\end{equation} 

The same argument also applies for the quark sector. However, in order to avoid the notorious tree-level flavor-changing neutral currents (FCNC), it is desirable to adopt natural flavor conservation~\cite{Glashow:1976nt,Paschos:1976ay}, that is, a $Z_2$ charge assignment is chosen such that right-handed quarks of the same charge couple only to a single Higgs doublet $\Phi_a$. Mathematically,
\begin{align}
{\cal L}_{\rm Yuk}^{\rm quarks} = Y^u_{ij} \bar Q_i \tilde{\Phi}_a u_{Rj} + Y^d_{ij} \bar Q_i \Phi_b d_{Rj} + \hc,
\end{align}
where $\tilde{\Phi}_a\equiv i\sigma_2\Phi^*_a$. After rotating the Higgs doublets using Eq.~\eqref{eq:rotation-Higgs}, we can write the Lagrangian as~\cite{Aoki:2009ha,Branco:2011iw}
\begin{align}
{\mathcal L}_{\rm Yuk}^{\rm quark} = \frac{m_{q_i}}{v} \left( \bar q_i q_ih + \xi^q_H \bar q_i q_iH -i\xi^q_A \bar q_i \gamma_5 q_i A \right)  - \left\{\frac{\sqrt{2}}{v} \bar u_i \left(\xi^d_A V_{ij}m_{d_j} P_R + \xi^u_A m_{u_i} V_{ij}P_L \right)d_j H^+ + \hc \right\},
\label{eq:q-Lag}
\end{align}
where $V$ denotes the Cabibbo-Kobayashi-Maskawa (CKM) mixing matrix and $P_{L,R}=\frac{1}{2}(1\mp\gamma_5)$ are projection operators. In the decoupling limit, we have $\xi^u_H=\xi^u_A$ and $\xi^d_H=-\xi^d_A$. It is then straightforward to see that, in the case of quarks coupling universally to $\{\Phi_1,\Phi_2\}$, $\xi^{u,d}_H=\{\tan\beta,-\cot\beta\}$. Similarly, in the case where only $\Phi_1$ couples to $d_{Rj}$ and $\Phi_2$ to $u_{Rj}$, as in the type-II 2HDM, we have $\xi^u_H=-1/\xi^d_H=-\cot\beta$.

Although the quark sector does not seem related to the neutrino sector, neutrino oscillation data will play an important role in determining whether the quarks couple to $\Phi_1$ or $\Phi_2$ for relatively light scalars. As we will see later in the next section, fitting the model into neutrino data will result in the allowed values of $\tbeta$. The very same parameter is also known to be constrained from rare $B$ decays, in particular the $B \to X_s\gamma$ transitions. From Ref.~\cite{Haller:2018nnx}, we can see that if quarks couple to $\Phi_2$ there exists a lower bound for $\tbeta$, albeit charged scalar mass dependent. For example, if $m_{H^+}=130$ GeV, $\tbeta\gtrsim 3$. This bound loosens to $\tbeta\gtrsim 1.3$ for $m_{H^+}=800$ GeV. Similarly, in the case where quarks couple to $\Phi_1$, this bound will translate to $\tbeta\lesssim0.33$ ($\tbeta\lesssim0.77$) for $m_{H^+}=130$ GeV (800 GeV). This suggests that, in the case where quarks couple universally to one Higgs doublet and the scalars are light, quarks are dictated to couple to $\Phi_1$ ($\Phi_2$) for low (large)~$\tbeta$. Although such $B \to X_s\gamma$ transitions only constrain the charged Higgs mass, whereas in the present paper we will mostly discuss the neutral Higgses, the bound is still relevant since the masses of exotic scalars cannot differ so greatly, or else there will be inconsistencies with the electroweak $\rho$ parameter constraint and perturbativity, see Refs.~\cite{Branco:2011iw,Gunion:2002zf}. 

In the case where $\Phi_1$ couples to $d_{Rj}$ and $\Phi_2$ to $u_{Rj}$, the bounds from $B \rightarrow X_s \gamma$ require that $m_{H^+} \gtrsim 600$ GeV for $\tan \beta\gtrsim2$. At the same time, $\tbeta$ is bounded from above by other rare $B$ decay processes. The strongest upper bound is provided by $B_s\to\mu^+\mu^-$, with $\tbeta\lesssim 17$ for $m_{H^+}=600$ GeV. The bound loosens to $\tbeta\lesssim22$ for $m_{H^+}=800$ GeV 
 If, instead, one considers the case where $\Phi_1$ couples to $u_{Rj}$ and $\Phi_2$ to $d_{Rj}$, the bounds from $B \rightarrow X_s \gamma$ would translate to $m_{H^+} \gtrsim 600$ GeV for $\tan \beta\lesssim 0.5$. The corresponding constraints from $B_s\to\mu^+\mu^-$ read $\tbeta\gtrsim 0.06$ (0.05) for $m_{H^+}=600$ (800) GeV. Improved measurements on $B \to {X_s} \gamma$ and $B_s \rightarrow \mu^+\mu^-$ will put more constraints on this allowed parameter space.

\allowdisplaybreaks

\section{Neutrino phenomenology}
\label{sec:neutrino}
In this section we study the neutrino phenomenology in our model. Neutrino masses are induced at the one-loop level by Feynman diagram shown in Fig.~\ref{fig:numass}. In addition to this, there is also another diagram whose internal particles are replaced by their charge conjugates, which is made possible by Majorana property of neutrinos. The sum of the two diagrams yields a symmetric neutrino mass matrix
\begin{equation}
	M_\nu = \kappa(\hat fM_{\ell}^{diag}\hat Y^T + \hat YM_{\ell}^{diag}\hat f^T), 
	\label{eq:nu-massmatrix}
\end{equation}
where $16\pi^2\kappa=\sin2\gamma\ln(M_{H_1^\pm}^2/M_{H_2^\pm}^2)$. Due to the fact that $\hat f$ contains a complex phase, the neutrino mass matrix is in general complex. It can be diagonalized by a unitary transformation, giving rise to real and positive eigenvalues 
\begin{equation}
	U^TM_\nu U = \text{diag.}(m_1,m_2,m_3),
	\label{eq:nu-diag}
\end{equation}
where $U$ is the Pontecorvo-Maki-Nakagawa-Sakata (PMNS) matrix. This transformation relates neutrinos in the flavor eigenbasis to those in the mass eigenbasis via
\begin{equation}
	\nu_i = U_{ia}\nu_a.
\end{equation} 
Here $i$ denotes family indices ($e,\mu,\tau$) and $a=1,2,3$. 

Note that, from Eq.~\eqref{eq:nu-diag}, the PMNS matrix $U$ will be parameterized by 3 mixing angles and 3 phases. The three neutrino mixing angles can be written as
\begin{equation}
	s_{12}^2 = \frac{|U_{e2}|^2}{1-|U_{e3}|^2},\quad
	s_{13}^2 = |U_{e3}|^2,\quad
	s_{23}^2 = \frac{|U_{\mu3}|^2}{1-|U_{e3}|^2},
	\label{eq:PMNS}
\end{equation}
with $s_{ij}$ ($c_{ij}$) standing for $\sin\theta_{ij}$ ($\cos\theta_{ij}$). One of the phases, known as the Dirac CP phase, is inferred from the Jarsklog invariance
\begin{equation}
	J = \text{Im}\left[U^{\phantom{\ast}}_{\mu3}U^{\ast}_{e3}U^{\phantom{\ast}}_{e2}U^{\ast}_{\mu2}\right].
	\label{eq:Jarlskog}
\end{equation} 
The other two phases, unmeasured so far, are called Majorana phases. 

Within this model, the neutrino mass matrix is expressed in terms of 6 free parameters. We will make a scan over 5 of them, which are two mixing angles of the charged lepton rotation matrices (chosen to be $\theta_1$ and $\alpha_3$), $r$ and $\theta$ of the $\hat f$ couplings, and $\tbeta$.  Then, by using Eqs.~\eqref{eq:nu-massmatrix}, \eqref{eq:nu-diag}, \eqref{eq:PMNS}, and \eqref{eq:Jarlskog}, we fit them to 5 neutrino observables, i.e., $s_{12}^2$, $s_{23}^2$, $s_{13}^2$, $J$, and $R\equiv\Delta m^2_{\sol}/\Delta m^2_{\atm}$, whose numerical values are presented in Table~\ref{tb:oscillation}. We demand that the aforementioned neutrino observables lie within their 2$\sigma$ ranges from their central values. To avoid the unknown Majorana phases, we will not use Eq.~\eqref{eq:nu-diag} directly, but instead diagonalize the Hermitian matrix $M_\nu^\dagger M_\nu$, through $U^\dagger M_\nu^\dagger M_\nu U = \text{diag.}\left(m_1^2,m_2^2,m_3^2\right)$. Once we find a fit, the absolute neutrino mass can be determined using $\kappa f_{e\mu}$ with the help of neutrino mass splittings.  

We first analyze the neutrino mass matrix in Eq.~\eqref{eq:nu-massmatrix} analytically. Note the mass matrix is approximately in the form
\begin{equation}
	(M_\nu)_{ij} \simeq \kappa[m_\tau(\hat f_{i3}\hat Y_{j3}+\hat f_{j3}\hat Y_{i3}) + m_\mu(\hat f_{i2}\hat Y_{j2}+\hat f_{j2}\hat Y_{i2})]\equiv (M_\nu^{(0)} + M_\nu^{(1)})_{ij}.
\end{equation}
Since the Yukawa coupling $\hat Y$ is hierarchical with $\hat Y_{ij}\sim m_{\ell_j}$, we see that the leading neutrino mass matrix is parametrically of order $m_\tau^2$ with a correction of order $m_\mu^2$. $M_\nu^{(0)}$ is a rank 2 matrix with a vanishing (3,3) element. This implies $\theta_{23}$ must lie in the first octant, see, e.g.,~\cite{He:2003ih,Babu:2013pma}. Moreover, since $M_\nu^{(0)}$ contains one vanishing eigenvalue, one would expect from the full mass matrix two sizable eigenvalues and one eigenvalue being suppressed by $m_\mu^2/m_\tau^2$. These observations, taken at face value, seem to favor invert ordering (IO) over normal ordering (NO) neutrino masses with $\theta_{23}<\pi/4$ . However, we note that  if the parameters of the model conspire to make $M_\nu^{(1)}$ becomes comparable to $M_\nu^{(0)}$, we could get a quasidegenerate neutrino mass compatible with both NO and IO. In this scenario, we could also accommodate $\theta_{23}>\pi/4$.

We next analyze the neutrino mass matrix numerically by performing a scan on the model parameter space. In our scan, we will consider both of NO and IO. For the solar mass splitting, we use the usual definition $\Delta m_{\sol}^2=m_2^2-m_1^2$, but for the atmospheric mass splitting, we adopt the convention used by Ref.~\cite{Esteban:2020cvm}, that is, $\Delta m^2_{\rm atm}=m_3^2-m_1^2$ for NO and $\Delta m^2_{\rm atm}=m_2^2-m_3^2$ for IO. Moreover, to reflect the current status of $\theta_{23}$, which can lie either in the first or second octant, we also consider its two global fit values. 
From our scan, we find that neutrino masses are quasidegenerate with $0.6\lesssim m_1/m_3\lesssim0.8$ for NO and $2\lesssim m_1/m_3\lesssim15$ for IO.

\begin{table}[t!]
\begin{center}
		\caption{Central values and the 1$\sigma$ range for neutrino oscillation parameters obtained from Ref.~\cite{Esteban:2020cvm} (see also~\cite{deSalas:2020pgw}). In our analysis, we also consider the case of which $\theta_{23}$ lies in the first octant.}
	\label{tb:oscillation}
\begin{tabular}{ccc}
\hline\hline
Parameters & Normal Ordering & Inverted Ordering \\
\hline
$s^2_{12}$ &$0.304\pm0.012$ &$0.304\pm0.013$\\
$s^2_{23}$ &$0.573^\pm0.018$ &$0.575\pm0.018$\\
$s^2_{13}$ &$0.02219\pm0.00063$ &$0.02238\pm0.00063$\\[.1cm]\hline
$\Delta m^2_{\sol}/10^{-5}~\text{eV}^2$ &$7.42\pm0.20$  &$7.42\pm0.20$  \\
$\Delta m^2_{\atm}/10^{-3}~\text{eV}^2$ &$2.517\pm0.027$  &$2.498\pm0.028$ \\
$\Delta m^2_{\sol}/\Delta m^2_{\atm}$ &$0.0295\pm0.0031$ &$0.0297\pm0.0034$\\[.1cm]\hline
$J_{CP}$ &$-0.0084^{+0.0127}_{-0.0143}$ &$-0.0327^{+0.0066}_{-0.0008}$\\
\hline\hline
\end{tabular}
\end{center}
\end{table}



\begin{figure}[h]
	\begin{center}
		\includegraphics[width=0.5\textwidth]{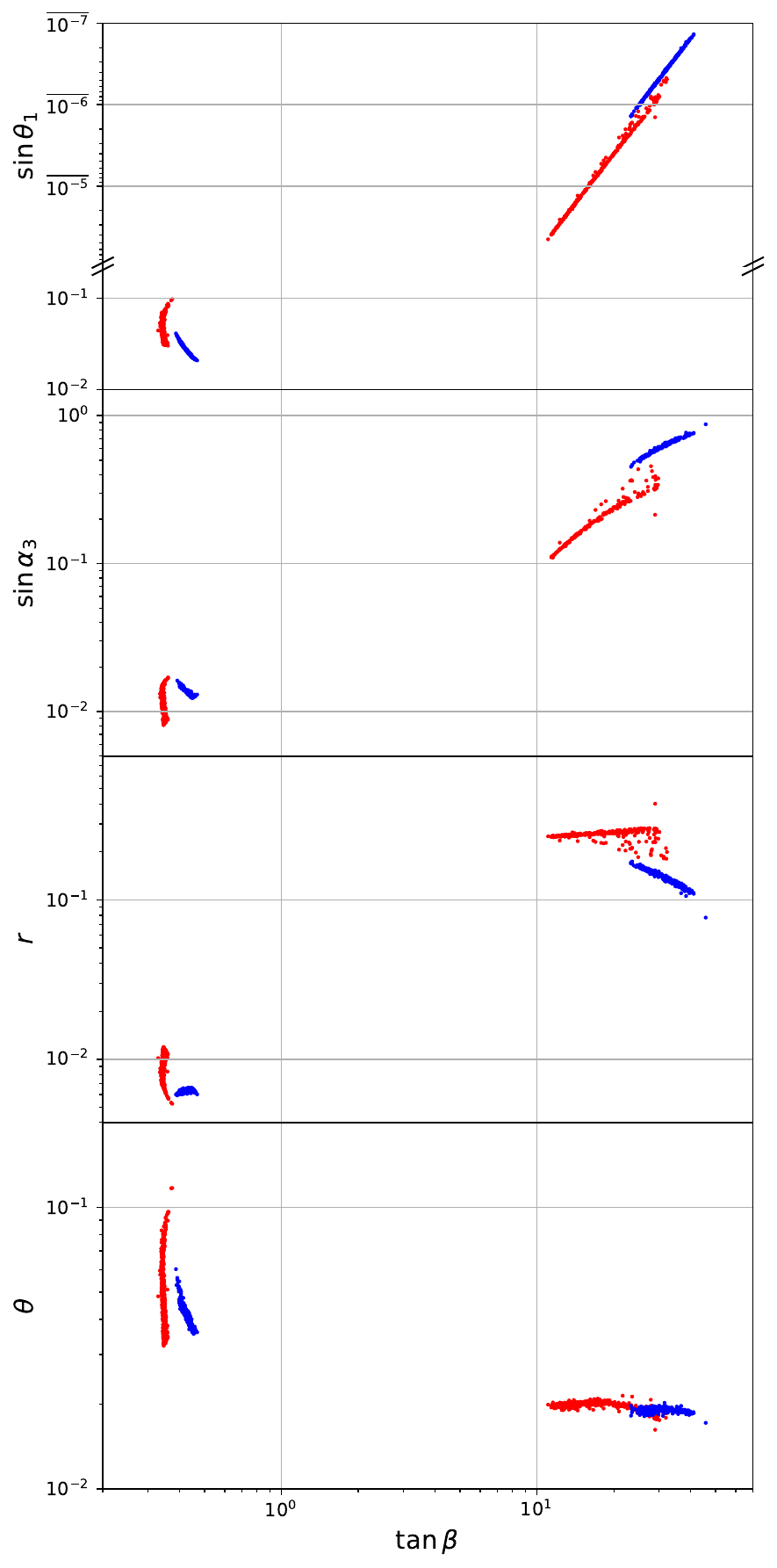}
		\caption{Parameter space consistent with neutrino oscillation parameters at 2$\sigma$ level and cosmology constraint on neutrino mass $\sum m_{\nu_i}<0.13$ eV. The parameter space is projected onto the $\sin(\theta_3)$ vs $\sin(\alpha_1)$, $\tbeta$, $r$ and $\theta$ plane (from top to bottom respectively.   Blue and red points correspond to IO neutrino masses. For red points, $\theta_{23}$ resides in the first octant. $\overline X = 1-X$}
		\label{fig:viableparam}
	\end{center}
\end{figure}

Since the neutrino masses are quasidegenerate, the parameter space is further constrained  by the bounds on neutrino masses.  The most stringent constraint is provided by cosmological observations. Assuming the standard $\Lambda$CDM model of the universe, Planck 2018 data places an upper bound $\sum m_\nu < 0.26$ eV at 95\% C.L.~\cite{Aghanim:2018eyx}. The bound can even be stronger if one combines Planck data with baryonic acoustic  oscillation (BAO) measurement, which puts bound on $\sum m_\nu < 0.13$ eV at 95\% C.L.~\cite{Aghanim:2018eyx}. In Fig.~\ref{fig:viableparam}, we present the viable parameter space consistent with neutrino oscillation parameters and the neutrino mass bound, i.e., $\sum m_\nu < 0.13$ eV. The viable parameter space is separated into two distinct regions, $\tbeta \lesssim 0.5$ and $\tbeta \gtrsim 10$. We will refer to them as the small and the large $\tbeta$, respectively. Note that the viable parameter space corresponds to the inverted neutrino masses ordering; the NO one is excluded by the Planck data, which is partly due to quasidegeneracy property of neutrino masses.  Moreover, both octants of  $\theta_{23}$ are allowed. In Fig.~\ref{fig:viableparam}, allowed regions corresponding to the first and second octants are depicted by red and blue points, respectively.

\begin{figure}[t!]
\begin{center}
\includegraphics[width=0.4\textwidth]{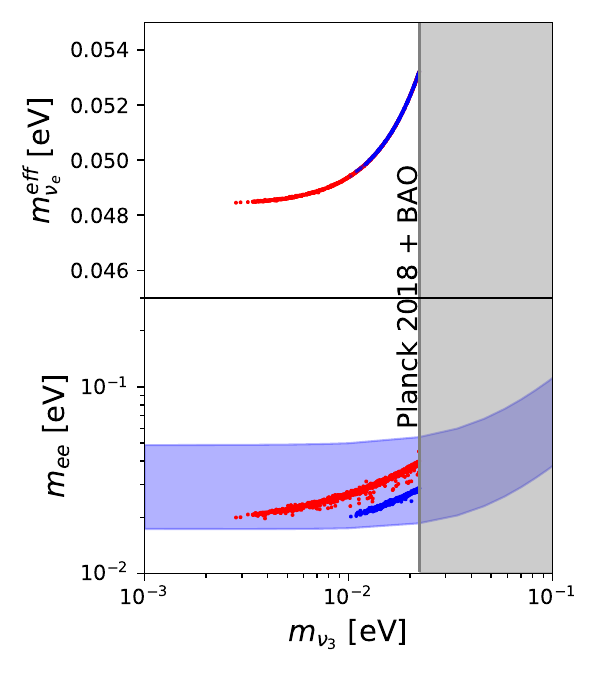}
\caption{Effective neutrino masses as a function of the lightest neutrino mass for beta decay experiments (top) and neutrinoless double beta decay experiment (bottom). On both plots, we show the Planck+BAO bound translated to $m_{\nu_3}$, $m_{\nu_3}\lesssim0.022$ eV, and on the bottom we also show the standard 95\% C.L. band for $m_{ee}$~\cite{ParticleDataGroup:2020ssz} for comparison.  For red (blue) points, $\theta_{23}$ resides in the first (second) octant.}
\label{fig:numassbound}
\end{center}
\end{figure}

We conclude this section by determining the effective neutrino masses, relevant for laboratory-based neutrino experiments, for our viable parameter space. First, we discuss the total neutrino masses, which is estimated in this model to be between 100 and 130 meV. This range is within reach of the Simon Observatory, which has projection of sensitivity up to 40 meV \cite{SimonsObservatory:2019qwx}. Second, experiments measuring the end point in beta decay spectrum use the effective electron neutrino mass, defined as $(m_{\nu_e}^{eff})^2= \sum_{i=1}^3 \left|U_{ei}^2\right|m_i^2$. For our viable parameter space, we have 48 meV $\lesssim m_{\nu_e}^{eff}\lesssim 56$ meV. This range is about 2 orders of magnitude below the current experimental limit of $m_{\nu_e}^{eff}<1100$ meV at 90\% C.L. reported by the KATRIN experiment~\cite{Aker:2019uuj}. Future experiments, like HOLMES \cite{Alpert:2014lfa}, will push this bound to 200 meV, which is still one order of magnitude higher than our estimate. Third, the neutrinoless double beta decay ($0\nu\beta\beta$) effective mass $m_{ee}$, defined as the (1,1) component of the neutrino mass matrix $m_{ee} = |\sum_{i=1}^3U_{ei}^2m_i|$,  lies in a range of 20 meV $\lesssim m_{ee}\lesssim40$ meV, see Fig.~\ref{fig:numassbound}. This is just below the best experimental limit $m_{ee}\lesssim 61-165$ meV at 90\% C.L. provided by the KamLAND-ZEN experiment~\cite{KamLAND-Zen:2016pfg}, which searches for $0\nu\beta\beta$ in $^{136}$Xe. The interval in upper bound reflects the uncertainty in the nuclear matrix elements used in extracting the limit. This range of values should be testable by phase-II LEGEND experiment \cite{LEGEND:2017cdu}, able to probe value between 13 and 29 meV.

\section{Constraints from lepton flavor violations}
\label{sec:lfv}
The lepton sector of the model contains tree-level FCNC. As a result, the parameter space of the model is strongly constrained by LFV measurements provided the extra Higgs bosons are not too heavy. In this section, we give analytic results for LFV observables. Some of the observables considered, such as the semileptonic decays, depend also on the quarks Yukawa couplings, albeit being flavor diagonal. To achieve this, we simply implement the  natural flavor conservation in the quark sector, hence forbidding tree-level FCNC of this sector.   
\subsection{LFV in leptonic decays}
\label{sec:lfvleptonic}

\begin{figure}[htp]
\begin{center}
\includegraphics[width=0.35\textwidth]{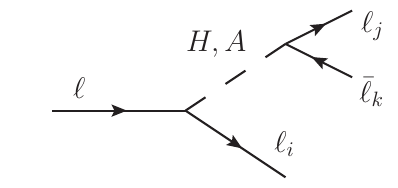}
\hspace{.2cm}
\includegraphics[width=0.35\textwidth]{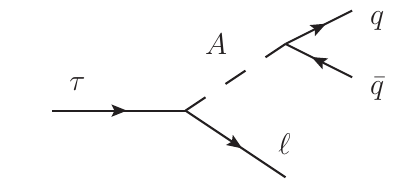}
\caption{Feynman diagram for leptonic decay $\ell \to \ell_i\ell_j\bar\ell_k$ (left) and semi-leptonic decay $\tau\to\ell P$ (right).}
\label{fig:leptonic}
\end{center}
\end{figure}

The LFV coupling matrix $\hat Y$ of the $H$ and $A$ lead to tree-level LFV decays of the lepton, $\ell \to \ell_i\ell_j\bar\ell_k$, via a Feynman diagram in Fig.~\ref{fig:leptonic}. This decay can be easily described in term of effective 4-fermion operators \cite{Celis:2014asa}
\begin{equation}
	\mathcal{L}_{eff} = c^{ijk}_{XY}(\bar \ell_{i}P_X\ell)(\bar \ell_jP_Y\ell_k),
\end{equation}
where $P_X,P_Y = P_L,P_R$ are the projection operators. The partial decay width for $\ell \to \ell_i\ell_j\bar\ell_k$ is then given by
\begin{equation}
	\Gamma_{\ell \to \ell_i\ell_j\bar\ell_k} = \frac{m_\ell^5}{512(1+\delta_{ij})\pi^3}\left(\frac{7}{768}|c^{ijk}_{XY}|^2+\frac{19}{768}|c^{jik}_{XY}|^2 -\frac{5}{384}\text{Re}\left[c^{ijk}_{XX}c^{jik\ast}_{XX}\right]\right),
\end{equation}
where we have ignored the final state lepton masses. In the above expression, the sum over projections is implicit. That is $|c^{ijk}_{XY}|^2=|c^{ijk}_{LL}|^2+|c^{ijk}_{RR}|^2+|c^{ijk}_{LR}|^2+|c^{ijk}_{RL}|^2$ and $|c^{ijk}_{XX}c^{jik\ast}_{XX}| = |c^{ijk}_{LL}c^{jik\ast}_{LL}| + |c^{ijk}_{RR}c^{jik\ast}_{RR}|$. In the case that all leptons in the final state are of the same family, the above expression reduces to
\begin{equation}
	\Gamma(\ell_i\to3\ell_j) = \frac{m_{\ell_i}^5}{1024\pi^3}\frac{\hat Y_{ji}^2\hat Y_{jj}^2}{384}\left(\frac{21}{m_H^4}+\frac{21}{m_A^4} - \frac{10}{m_H^2m_A^2}\right) + \mathcal{O}\left(\frac{m_{\ell_j}^2}{m_{\ell_i}^2}\right).
\end{equation} 
Current experimental limits are $Br(\mu\to3e)<1.0\times10^{-12}$, $Br(\tau\to3e)<2.7\times10^{-8}$  and $Br(\tau\to3\mu)<2.1\times10^{-9}$ at 90\% C.L.~\cite{PDG:2020ssz}. On the viable parameter space of the model, these constraints from tree-level leptonic LFV decays are much weaker compared to radiative LFV decays.

\subsection{LFV in semileptonic decays}

The LFV coupling of the $A$ also lead to a semi-leptonic decays $\tau\to\ell P$, where $P$ is a neutral pseudoscalar meson. The partial decay width, ignoring the final state lepton mass, is given by~\cite{Celis:2014asa,Primulando:2016eod}
\begin{align}
	\Gamma(\tau\to\ell \pi) &= \frac{(m_\tau^2-m_\pi^2)^2}{256\pi m_\tau^3}\frac{m_\pi^4f_\pi^2}{v^2m_A^4}\left(\xi^u_A-\xi^d_A\right)^2(\hat Y_{\ell\tau}^2+\hat Y_{\tau\ell}^2),\\
	\Gamma(\tau\to\ell\eta^{(\prime)}) &=  \frac{(m_\tau^2-m_{\eta^{(\prime)}}^2)^2}{256\pi m_\tau^3}\frac{1}{v^2m_A^4}\left((\xi^u_A+\xi^d_A)h^q_{\eta^{(\prime)}}+2\xi^d_Ah^s_{\eta^{(\prime)}}\right)^2(\hat Y_{\ell\tau}^2+\hat Y_{\tau\ell}^2),
\end{align}
where $\xi^q_A$ is the coefficient defined just right after Eq.~\eqref{eq:q-Lag} and 
$f_\pi\simeq (130.2\pm0.8)$ MeV~\cite{Aoki:2019cca} is the pion decay constant and 
\begin{equation}
	h^q_{\eta} = h^q_{\eta'} =0.001\pm0.003 \text{ GeV}^3,\quad
	h^s_{\eta} = -0.055\pm0.003 \text{ GeV}^3,\quad
	h^s_{\eta^{\prime}} = 0.068\pm0.005 \text{ GeV}^3,
\end{equation}
are the hadronic matrix elements~\cite{Beneke:2002jn}. 
Since the decays proceed only via the pseudoscalar $A$, one can use limits on the branching ratio to constrain the LFV coupling $\hat Y$ as a function of $m_A$. We find that the decay $\tau\to\ell\eta$ gives the strongest bounds. 
Using $Br(\tau\to e\eta)<9.2\times10^{-8}$ and $Br(\tau\to\mu\eta)<6.5\times10^{-8}$~\cite{PDG:2020ssz}, the upper limits are
\begin{align}
	|\hat Y_{e\tau}| \lesssim \frac{0.97}{|\xi^d_A|}\left(\frac{m_A}{130 \text{ GeV}}\right)^2,\\
	|\hat Y_{\mu\tau}| \lesssim \frac{0.81}{|\xi^d_A|}\left(\frac{m_A}{130 \text{ GeV}}\right)^2.
\end{align} 
where we have dropped the suppressed coupling $\hat Y_{\tau\ell}$ and the hadronic matrix element $h^q_\eta$. Notice that the bounds only depend on the strange quark coupling, $|\xi^d_A|$. As an illustrative example of the above bounds, consider the scenario where $m_A$ is around the weak scale and $0.3\lesssim\tbeta\lesssim50$. If strange quark couples to $\Phi_2$, we have $|\xi^d_A|=\cot\beta$. In this case, the strongest upper bound on the LFV coupling $|\hat Y_{e\tau}|$ and $|\hat Y_{\mu\tau}|$ is around $\mathcal{O}(10^{-1})$. If instead, strange quark couples to $\Phi_1$ so that $|\xi^d_A|=\tbeta$, the strongest upper bound becomes $\mathcal{O}(10^{-2})$. In both cases , the strongest upper bound is well above the typical value of the couplings in our viable parameter space.

\subsection{LFV in radiative decays}

Equation \eqref{eq:leptonyuk} can also induce radiative LFV decays. For a general $\ell \to \ell'\gamma^*$ transition, its effective Lagrangian takes the form of (see, e.g., \cite{kuno:1999jp})
\begin{equation}
	\mathcal{L}_{\ell\to\ell'\gamma^*} = \frac{em_\ell}{4\pi^2} \bar \ell' i\sigma^{\mu\nu}q_\nu(c_L P_L + c_R P_R)\ell  A_\mu
	+ e \bar \ell' \gamma^\mu (a_L P_L + a_R P_R) \ell A^{\nu} \left(q^2 g_{\mu\nu} - q_\mu q_\nu \right)
	+ \hc,
	\label{mu-e-gamma}
\end{equation}
where $q\equiv p_\ell-p_{\ell'}$ is the momentum transfer of the photon. Note the vanishing of the non-dipole term when the photon is on-shell. The form of Eq.~\eqref{mu-e-gamma} also guarantees the fulfillment of Ward's identity. For $\ell\to\ell'\gamma$ process, its decay width is given by
\begin{equation}
	\Gamma(\ell\to\ell'\gamma) = \frac{\alpha_{em} m_\ell^5}{64\pi^4}\left(|c_L|^2+|c_R|^2\right),
\end{equation}
where~$\alpha_{em}\equiv e^2/4\pi$ is the QED fine structure constant. At one loop, such decay proceeds through penguin diagrams mediated by neutral and charged scalars. Its contributions to those Wilson's coefficients are found to be~\cite{Hisano:1995cp,Arganda:2005ji}
\begin{align}
	c_L^{\rm 1-loop} &=  -\frac{\Y_{a\ell}\Y_{a\ell'}}{24} \sum_{\phi=H,A,H^+} \frac{(-1)^{Q_\phi}}{m_\phi^2} + \frac{\Y_{\ell a}\Y_{a\ell'}}{8} \left(\frac{m_{a}}{m_\ell}\right) \sum_{\phi=H,A} 
	\frac{(-1)^{CP}}{m_\phi^{2}} \left(3 + 2\ln \frac{m_{a}^2}{m_\phi^2} \right), \label{cL-1}\\
c_R^{\rm 1-loop} &=  -\frac{\Y_{\ell a}\Y_{\ell' a}}{24} \sum_{\phi=H,A} \frac{1}{m_\phi^2} + \frac{\Y_{a\ell}\Y_{\ell' a}}{8} \left(\frac{m_{a}}{m_\ell}\right) \sum_{\phi=H,A} 
	\frac{(-1)^{CP}}{m_\phi^{2}} \left(3 + 2\ln \frac{m_{a}^2}{m_\phi^2} \right)\label{cR-1},
\end{align}
where $Q_\phi$ and $CP$ denote the electric charge and charge-parity eigenvalue of the respected scalar. The first term in each equation above denotes the case where the chirality flip occurs in the external leg of the decaying lepton, while the second term denotes the flip in the internal leptons.   

The 2--loop contributions to $c_{L,R}$ arise via Barr-Zee diagrams. This type of diagrams, especially those containing top-quark or $W$ boson loop, may be equally or even more important than those of the 1--loop, due to the 2--loop suffering less chirality suppression and containing fewer $\hat Y_{ab}$, which is typically smaller than the top-Yukawa or $SU(2)$ gauge couplings~\cite{Bjorken:1977vt,Barr:1990vd,Chang:1993kw,Davidson:2010xv}. In the present case, we adopt the decoupling limit, so the $W$--loop vanishes. This makes top loop the sole contribution to $c_{L,R}^{\rm 2-loop}$, i.e.,
\begin{align}
	c_{L}^{\rm 2-loop} &=  -\frac{\alpha_{em}}{\pi} \frac{ \hat Y_{\ell\ell'}}{v m_\ell} \xi^u_H \left(\frac{2}{3}\left[f(z_{tH}) - g(z_{tA})\right] + \frac{(1-4s_W^2)(1-8/3\,s_W^2)}{16s_W^2c_W^2}\left[\tilde f(z_{tH},z_{tZ}) - \tilde g(z_{tA},z_{tZ})\right]\right), 
\label{cL-2}\\
	c_{R}^{\rm 2-loop} &=  -\frac{\alpha_{em}}{\pi} \frac{ \hat Y_{\ell'\ell}}{v m_\ell} \xi^u_H\left(\frac{2}{3}\left[f(z_{tH}) + g(z_{tA})\right] + \frac{(1-4s_W^2)(1-8/3\,s_W^2)}{16s_W^2c_W^2}\left[\tilde f(z_{tH},z_{tZ}) + \tilde g(z_{tA},z_{tZ})\right]\right), 
\label{cR-2}
\end{align}
where $z_{ab}\equiv m_a^2/m_b^2$. The first term in brackets is the photon-exchange contribution with loop functions expressed as
\begin{align}
	&f(z) = \frac{z}{2}\int_0^1\text{d}x\, \frac{1-2x(1-x)}{x(1-x)-z}\ln\frac{x(1-x)}{z}; \quad
	g(z) =\frac{z}{2}\int_0^1\text{d}x\, \frac{1}{x(1-x)-z}\ln\frac{x(1-x)}{z},
\end{align}
whereas the second term denotes the $Z$ contribution with functions  defined as $\tilde h(x,y)=\left(xh(y)-yh(x)\right)/(x-y)$. Due to the factor of $1-4s_W^2$, one should quickly recognize that the $Z$ contribution is much suppressed compared to that of photon. 

For the photon exchange, the functions $f(z)$ and $g(z)$ differ only on $1-2x(1-x)$, which  is a positive definite function. Therefore, the two functions will always come with the same sign. In the case of $c_L^{\rm 2-loop}$, this leads to a destructive interference between $H$ and $A$ loops. For a certain mass set up, it could even lead to a complete cancelation, leaving us with the suppressed $Z$ contribution. Along with the fact that $|\hat Y_{\ell \ell'}/\hat Y_{\ell'\ell}|\sim m_{\ell'}/m_\ell$, we have $c_L^{\rm 2-loop} \ll c_R^{\rm 2-loop}$. 

For our viable parameter space obtained in Sec.~\ref{sec:neutrino}, see Fig.~\ref{fig:viableparam}, $\mu\to e\gamma$ measurement places the strongest constraint on it. If the extra Higgs boson are not too heavy, $\mu\to e\gamma$ decay generically rules out the part of parameter space with small $\tbeta$. This is because the LFV coupling $\hat{Y}_{e\mu}$ for the low $\tbeta$ region is enhanced while its counterpart in the high $\tbeta$ region is suppressed. 

\begin{figure}[htp]
\begin{center}
\includegraphics[width=0.5\textwidth]{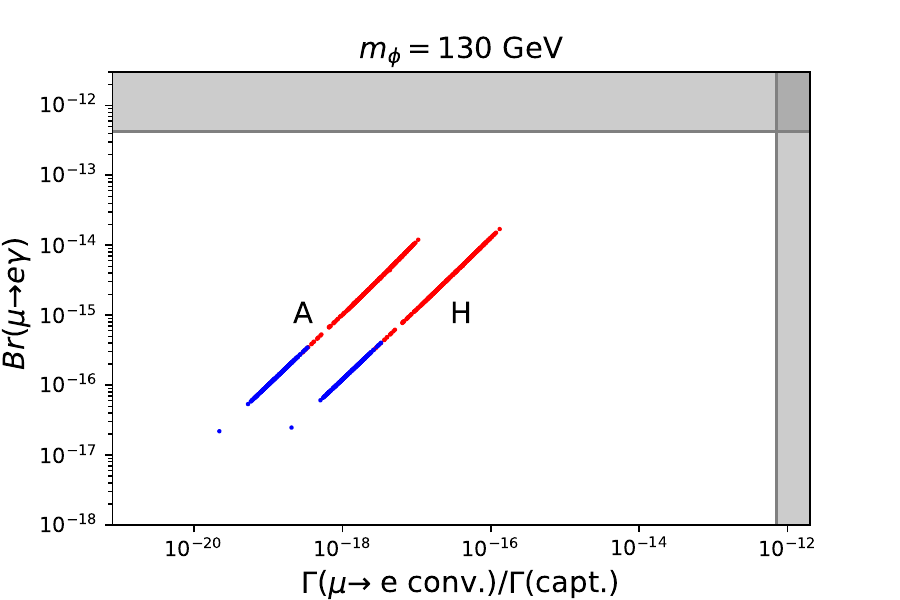}
\caption{The rates for $\mu\to e \gamma$ and $\mu\to e$ conversion for parameter space identified by neutrino data. Here, $m_\phi$ stands for the smallest of $m_H$ and $m_A$.  Red (blue) points correspond to $\theta_{23}$ being in the first (second) quadrant. The gray regions are experimentally excluded.}
\label{fig:lowenergylfv}
\end{center}
\end{figure}

For concreteness, we consider benchmark scenario where only one neutral scalar is light with $m_\phi=130$ GeV, where $\phi=H/A$, while other Higgs are heavy. This choice of benchmark is relevant for collider search to be discussed in Sec.~\ref{sec:heavyhiggs}. To be specific, the heavier Higgs bosons are taken to be 650 GeV. The mass splitting in the Higgs sector chosen in this benchmark is consistent with perturbativity and the electroweak $\rho$-parameter constraints on the scalar quartic couplings~\cite{Branco:2011iw}. Moreover, since our $\tbeta$ are large, constraints from $B$ physics discussed in Sec.~\ref{sec:model} dictate that top quark couples to $\Phi_2$. The constraints from $\mu\to e\gamma$ transition for our benchmark scenario is shown in Fig.~\ref{fig:lowenergylfv}. Again, red (blue) points correspond to $\theta_{23}$ being in the first (second) octant.  
Note that the branching ratio $\mu\to e\gamma$ decreases with increasing  $m_\phi$, as expected.

\subsection{$\mu\to e$ conversion in nuclei}
The process of $\mu\to e$ conversion in atomic nuclei also provides a stringent constraint on the LFV coupling, in particular $\hat Y_{e\mu}$ and $\hat Y_{\mu e}$. The formalism of such process has been discussed intensively in Refs.~\cite{Kitano:2002mt,kuno:1999jp}. We will focus on the coherent transition, where the nucleus final state does not change from its initial state during the conversion. Because of that, the effective Lagrangian at the quark level takes the form of
\begin{equation}
\mathcal{L}_{\rm eff} =  \frac{em_\mu}{8\pi^2} c_L \bar e \sigma^{\alpha\beta} P_L \mu  F_{\alpha\beta} - \frac{1}{2} \sum_q \left[ (g^q_{LS} \bar e P_R \mu )(\bar{q}q) + (g_{LV}^q \overline{e} \gamma^\alpha P_L \mu )(\bar{q}\gamma_\alpha q) \right]  + (L \leftrightarrow R) + \hc
\label{mu-e-conv}
\end{equation}
This quark-level Lagrangian get converted into the nucleon-level one by evaluating the corresponding nuclear matrix elements. That is, $\left<N|\bar q\Gamma_K q|N\right>\equiv G_K^{(q,N)} \bar{N}\,\Gamma_K N$ with $\Gamma_K=\{1,\gamma_\alpha\}$ and $N=p,n$. Other types of operators such as  $\bar q\gamma_5 q$, $\bar q\gamma_\alpha\gamma_5 q$, and  $\bar q\sigma_{\alpha\beta}q$  have no contribution to the coherent process, hence they are omitted from  Eq.~\eqref{mu-e-conv}.

The couplings of each operator can be directly determined. For vector operators, which are induced by the monopole term of Eq.~\eqref{mu-e-gamma}, they are  given by $g^q_{LV,RV} = -8\pi\alpha_{em}Q_qa_{L,R}$, with $Q_q$ symbolizing the quark electric charge whereas $a_{L,R}$ being loop functions \cite{Hisano:1995cp,Arganda:2005ji}
\begin{align}
a_L = -\frac{1}{144\pi^2}\Y_{\mu a} \Y_{ea} \sum_{\phi=H,A} \frac{1}{m_\phi^2}\left(4+3\ln \frac{m_a^2}{m_\phi^2} \right), \quad 
a_R = -\frac{1}{144\pi^2}\Y_{a\mu} \Y_{ae} \left[\sum_{\phi=H,A} \frac{1}{m_\phi^2}\left(4+3\ln \frac{m_a^2}{m_\phi^2} \right) -\frac{1}{m_{H^+}^2} \right].
\end{align}
The scalar operators arise via tree-level exchange of $H$, which gives
\begin{align}
g_{LS}^q = -\left(\frac{2}{m_H^2}\right)\left(\frac{m_q}{v}\right)\Y_{e\mu}\xi^q_H, \quad g_{RS}^q = -\left(\frac{2}{m_H^2}\right)\left(\frac{m_q}{v}\right)\Y_{\mu e}\xi^q_H
\end{align}

Now the conversion rate can be calculated by using the formula given in Ref.~\cite{Kitano:2002mt} (see also Ref.~\cite{Harnik:2012pb}), that is, 
\begin{align}
	\Gamma(\mu\to e\text{ conv.}) &= \left|-\frac{e}{16\pi^2}c_RD + \tilde{g}^{(p)}_{LS} S^{(p)} + \tilde{g}^{(n)}_{LS} S^{(n)} + \tilde{g}^{(p)}_{LV}V^{(p)} \right|^2 \nn\\
	&\quad +  \left|-\frac{e}{16\pi^2}c_LD + \tilde{g}^{(p)}_{RS} S^{(p)} + \tilde{g}^{(n)}_{RS} S^{(n)} + \tilde{g}^{(p)}_{RV}V^{(p)} \right|^2,
	\label{mu-e-convrate}
\end{align}
where $D$, $S^{(p,n)}$, and $V^{(p)}$ are the corresponding overlap integrals for each operator. Their definitions and numerical values (in the unit of $m_\mu^{5/2}$) for various nuclei have been tabulated in \cite{Kitano:2002mt}. The coefficients $\tilde g$'s stand for effective couplings to nucleon $N$, which are defined as
\begin{align}
\tilde g^{(N)}_{LK,RK} = \sum_q G^{(q,N)}_K g^q_{LK,RK}.
\end{align}

One should note that there is a subtle difference  between the vector and scalar operators. The vector operators characterize the difference between the number of quarks and antiquarks. Thus the summation runs over the valence quarks with $G^{(u,p)}_V=G^{(d,n)}_V=2$ and $G^{(u,n)}_V=G^{(d,p)}_V=1$. This leads to the vanishing effective neutron couplings $\tilde g^{(n)}_{LV,RV}=0$. The scalar operators, on the other hand, quantify each quark contribution to nucleon mass via $G^{(q,N)}_S = f^{(q,N)}m_N/m_q$. Therefore, the sum runs for all quark flavors, including those of heavy flavors ($Q=c,b,t$) mediating the gluon exchange.  For light quarks $q=u,d,s$, the parameters $f ^{(q,N)}$ have been determined by lattice calculations and are presented in Table~\ref{tab:formfactor}, while for heavy quarks, it is determined through 
\begin{align}
f^{(Q,N)}=\frac{2}{27}\left(1-\sum_q f^{(q,N)}\right).
\end{align}
It should be noted that in both operators $G^{(u,p)}_{S,V}=G^{(d,n)}_{S,V}$ and $G^{(u,n)}_{S,V}=G^{(d,p)}_{S,V}$, reflecting the isospin invariance of the nucleon.

For our purpose, we seek the strongest constraint, coming from $\mu\to e$ conversion on gold nucleus. The branching fraction of the rate is $\Gamma(\mu\to e$ conv.)/$\Gamma$(captured) $< 7\times10^{-13}$ at 90\% C.L.~\cite{Bertl:2006up}. The overlap integrals are given by $D=0.189m_\mu^{5/2}$, $S^{(p)}=0.0614m_\mu^{5/2}$ and $S^{(n)} = 0.0918m_\mu^{5/2}$. The present upper bound of $\mu\to e$ conversion rate in gold nucleus is given by $\Gamma({\rm captured}) < 13.07\times10^6$ s$^{-1}$~\cite{Kitano:2002mt}. The  rate for our benchmarks is shown in Fig.~\ref{fig:lowenergylfv}.

\begin{center}
\begin{table}[t]
\caption{Numerical values for $f^{(q,N)}$. Note that $Q$ runs for all heavy quarks, $c,b,t$.}
\label{tab:formfactor}
\begin{tabular}{|c|p{2.2cm}p{2.2cm}p{2.2cm}p{2.2cm}|}
\hline\hline
Nucleon & $f^{(u,N)}$~\cite{Bishara:2015cha} & $f^{(d,N)}$~\cite{Bishara:2015cha} &$f^{(s,N)}$~\cite{Junnarkar:2013ac} &$f^{(Q,N)}$~\cite{Harnik:2012pb} \\\hline
$p$ & 0.018$\pm$0.005& 0.034$\pm$0.011 & 0.043$\pm$0.011& 0.067$\pm$0.001\\\hline
$n$ & 0.016$\pm$0.005& 0.038$\pm$0.011 &0.043$\pm$0.011 & 0.067$\pm$0.001\\\hline\hline
\end{tabular}
\end{table}
\end{center}

\subsection{Anomalous magnetic dipole moment of muon}
In addition to LFV processes, the Yukawa couplings $\hat Y$ also contribute to the anomalous magnetic dipole moment of the muon, $a_\mu = (g_\mu-2)/2$. The SM has predicted $a_\mu^{\rm SM}=116591810(43)\times10^{-11}$~\cite{Aoyama:2020ynm}, which is disfavored at 4.2$\sigma$ by recently updated experimental measurement $a_\mu^{\rm exp.}=116592061(41)\times10^{-11}$~\cite{Muong-2:2021ojo}. (Others claim that their version of lattice QCD calculation of hadronic contribution to $a_\mu$ can lower the discrepancy  to 2.4$\sigma$~\cite{Borsanyi:2020mff}.) Nevertheless, such discrepancy can be attributed to a possible presence of new physics, which, in present model, arises via scalar exchanges. The new contribution is given by~\cite{Hisano:1995cp,Davidson:2010xv}
\begin{align}
	\delta a_\mu^{\rm NP} &= -\frac{m_\mu^2}{2\pi^2}\,\left.\text{Re}\left(c_L + c_R\right)\right|_{\ell=\ell'=\mu},
\end{align}
where $c_L$ and $c_R$ are the Wilson's coefficients given in Eqs.~\eqref{cL-1}~--~\eqref{cR-2} with $\ell=\ell'=\mu$.
Note that the contribution of the charged Higgs is negative, while the contribution of the neutral Higgses can take either sign, depending on masses and couplings. The new physics contribution in our scenario yields $a_\mu\sim \mathcal{O}(10^{-11})$ or smaller, deemed insufficient to account for the discrepancy.

\section{Neutral Higgs collider phenomenology}
\label{sec:heavyhiggs}
In this section, we study the LHC phenomenology of the new neutral scalars, i.e., $H$ and $A$. 
For light $H$ and $A$ within the collider reach, the parameter space consistent with both neutrino data and LFV constraints corresponds to $\tbeta\gtrsim 10$. 
Firstly, we will discuss the case of the quarks coupled universally to one of the scalar doublet. Since only high value of $\tbeta$ is allowed, constraints from $B$ physics~\cite{Haller:2018nnx}, such as $B_d \rightarrow \mu\mu$,  restrict the case where quarks couple only to $\Phi_2$ provided that $m_{H^+} \lesssim 1$ TeV.\footnote{The flavor phenomenology of quarks that couple universally to $\Phi_1$ with high value of $\tan \beta$ would be the same with the type-I 2HDM \cite{Branco:2011iw} with low $\tan\beta$, hence excluded by the $B$ physics constraints.} 
Since we are working in the decoupling limit, the main production channels for both the $H$ and the $A$ are gluon fusion via the top quark loop. They can also be produced in the association with the top pair or the bottom pair, but we found that these channels are insignificant compared with gluon fusion. 

The $H$ production cross-section can be calculated from the would-be SM Higgs boson by a rescaling the quark Yukawa couplings with $\xi^u_H$. In our analysis, we use the would-be SM Higgs boson cross-section provided in Ref.~\cite{Heinemeyer:2013tqa}. On the other hand, the pesudoscalar-gluon fusion cross-sections is larger compared to that of $H$ with the same mass, with a loop factor of
\begin{equation}
	\frac{\sigma^{A}_{ggF}}{\sigma^H_{ggF}} \simeq \frac{F(x)}{1+(1-x)F(x)},
\end{equation} 
where $x = 4m_t^2/m_{H(A)}^2$ and
\begin{equation}
	F(x) = \left\{\begin{aligned}&\left(\sin^{-1}\sqrt{1/x}\right)^2,\hspace{2.65cm}x\ge1,\\
	&-\frac{1}{4}\left[\ln\left(\frac{1+\sqrt{1-x}}{1-\sqrt{1-x}}\right)-i\pi\right]^2,\quad x<1.\end{aligned}\right.
\end{equation}

\begin{figure}[htp]
	\begin{center}
		\includegraphics[width=0.5\textwidth]{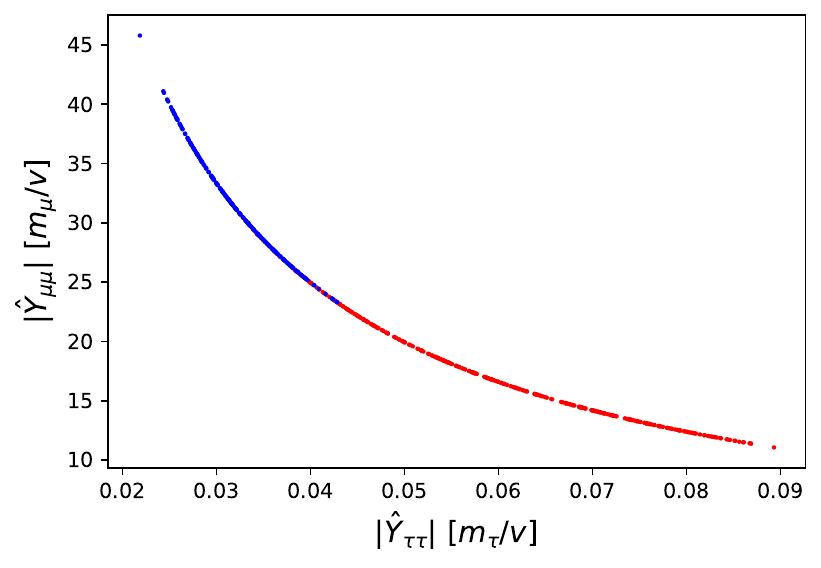}
		\caption{The flavor diagonal couplings $\hat Y_{\ell\ell}$ in the unit of $m_\ell/v$. The red (blue) points corresponds to $\theta_{23}$ lies in the first (second) octant.}
		\label{fig:yhat}
	\end{center}
\end{figure}

Since $\tbeta$ is large over the parameter space of interest,  both the $H$ and the $A$ production cross-sections and decay rates to quarks are suppressed. However, the Yukawa couplings of $H$ and $A$ to the leptons can be enhanced. The couplings $\hat Y_{\mu\mu}$ and $\hat Y_{\tau \tau}$ in the case of allowed parameter space are shown in Fig.~\ref{fig:yhat}. From the figure we see that $\hat Y_{\mu\mu}$ is enhanced compared with the SM Higgs Yukawa coupling. The value of $\hat Y_{ee}$ also gets comparable enhancements compared with its would be SM-value. Moreover, since this region is safe from the LFV constraints, in this region the LFV couplings ($\hat Y_{e \mu}$, $\hat Y_{e \tau}$, $\hat Y_{\mu \tau}$) are relatively small. As a result, the relevant LHC signature $H/A\to\mu^+\mu^-$ is considerably large if the scalars are within the LHC reach. On the other hand, while $\hat Y_{ee}$ gets some enhancements, the braching fractions of $H/A\to e^+e^-$ are still relatively small, hence less relevant in constraining the model.

\begin{figure}[htp]
\begin{center}
\includegraphics[width=0.45\textwidth]{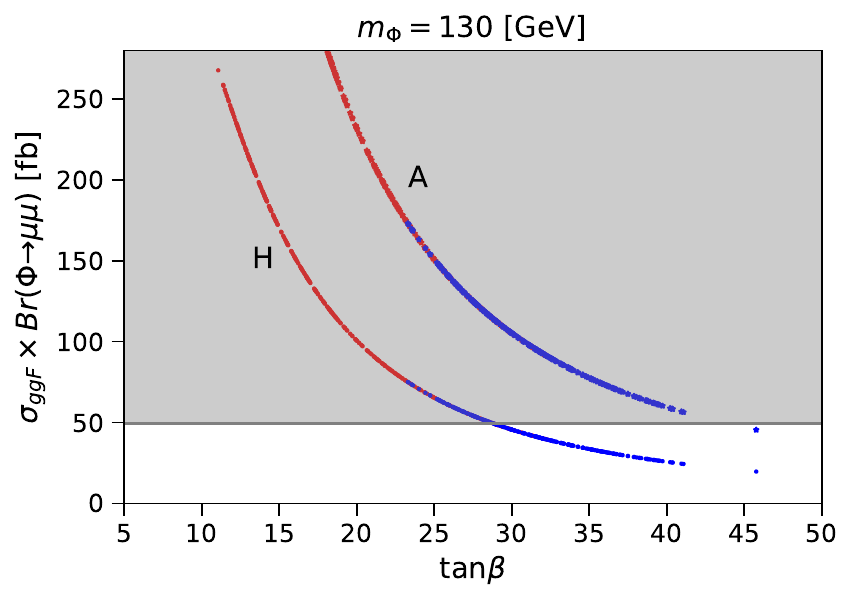}
\includegraphics[width=0.45\textwidth]{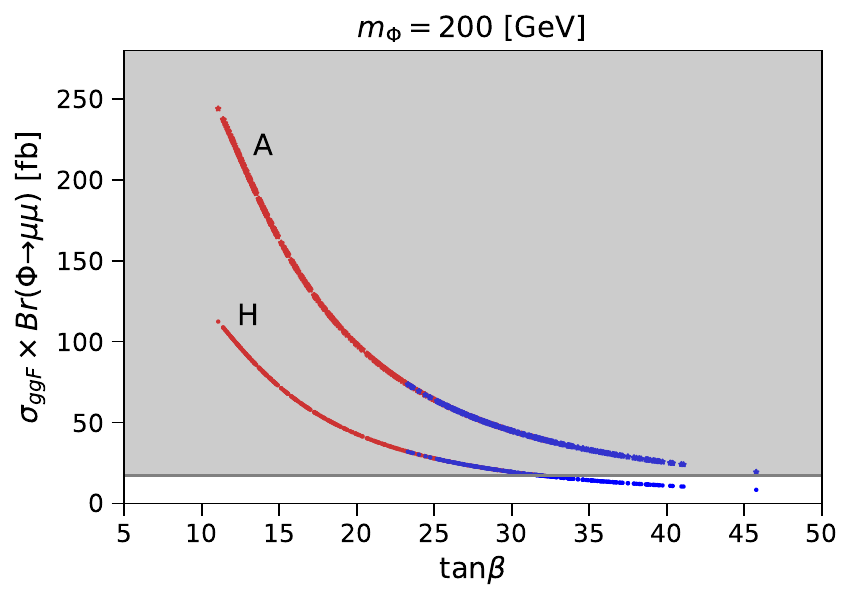}
\caption{The gluon fusion production cross-section times branching ratio into $\mu^+\mu^-$ as a function of $\tbeta$, in the case quarks couple to $\Phi_2$, for a light neutral Higgs boson $m_\phi=130$ GeV and $m_\phi=200$ GeV. The gray region is ruled out by the CMS search~\cite{CMS:2019mij}. For red (blue) points, $\theta_{23}$ resides in the first (second) octant.}
\label{fig:higgstomumu}
\end{center}
\end{figure}

At the 13-TeV LHC, CMS has searched for a supersymmetric Higgs boson decaying to $\mu^+\mu^-$. Based on 35.9 fb$^{-1}$ data, CMS reports the 95\% upper limit on the gluon fusion production cross-section ($\sigma$) times branching ratio (BR) into $\mu^+\mu^-$, ranging from 44.7 fb at $m_H =130$ GeV to 17.5 fb at $m_H=200$ GeV~\cite{CMS:2019mij}. We translate the CMS bounds into the constraints on our parameter space in the case $m_{H/A}=130$ GeV and $m_{H/A}=200$ GeV, see Fig.~\ref{fig:higgstomumu}. In the two plots we independently show bounds for each $H$ and $A$. One can see that if the pseudoscalar $A$ is the lightest, the LHC bounds become more stringent due to larger cross section. In either case, the lower value of $\tan \beta$, corresponding to the red points (see Fig. \ref{fig:viableparam}), is disfavored. As shown in Fig. \ref{fig:yhat}, the blue points have larger values of $\hat Y_{\mu\mu}$ hence larger branching fraction to the $\mu^+ \mu^-$. However, the production cross-section is suppressed by $1/\tbeta^2$, which weakens the collider bound. The LHC bounds start to be irrelevant once other decay channels, such as $A \rightarrow h Z$ and $H \rightarrow hh$, open up and dominate the branching fractions. 

For light enough $H$ or $A$, collider bounds are stronger than the LFV bounds with only little parts of allowed parameter space left. We expect that these bounds will get stronger once the full LHC Run II data are  analyzed with possibility of ruling out the whole region at the end of LHC runs.

Since only large $\tan\beta$ allowed from LFV bounds, there is only one more additional scenario for light scalar that satisfies the bounds from rare $B$ decays. This scenario is when $\Phi_1$ couples to $d_{Rj}$ and $\Phi_2$ to $u_{Rj}$, as in type-II 2HDM. In this scenario, the Yukawa coupling to the down-type quark is enhanced by a factor of $\tan\beta$ while the coupling to the up-type quark is reduced by the same factor. Therefore, while $\hat Y_{\mu\mu}$ is increased to the would-be SM coupling, the value of the Yukawa coupling is still smaller than the value of $Y_{b \bar b}$. The branching fractions of $H$ and $A$ are then dominated by the decay to $b \bar b$. For low value of $m_H$, CDF and D0 presented combined analysis for this channel for 90 GeV$ < m_H < 300 $ GeV \cite{CDF:2012btt}. Their bounds are rather weak, e.g. $\tan \beta \lesssim 40$ for $m_H = 130$ GeV. CMS and ATLAS also performed some analysis in this search channel \cite{ATLAS:2019tpq, CMS:2018hir}. However, the collaborations only show the bounds for $m_H \gtrsim 300$ GeV. An analysis at lower mass will be interesting on probing this scenario of quark-scalar couplings.

\section{Conclusions and discussions}
\label{sec:conclusion}
We have analyzed the Zee model for radiative neutrino mass generation with a family dependent discrete $Z_2$ symmetry imposed on the left-handed lepton doublets. The $Z_2$ symmetry dictates the resulting neutrino masses, consistent with oscillation data, to be quasidegenerate with $\tbeta$ predicted to be in a restrictive range---$\tbeta\lesssim0.5$ (small $\tbeta$) and $\tbeta\gtrsim10$ (large $\tbeta$), see Fig.~\ref{fig:viableparam}. 
Combining oscillation data with constraints from cosmology on the sum of neutrino masses, we found that the viable parameter space  supports only the inverted neutrino masses ordering (IO).

The viable parameter space of the model considered here is well placed to be probed by both the ongoing and the planned experiments. On the neutrino mass front, see Fig~\ref{fig:numassbound}, our scenario predicts the sum of the three active neutrino mass, $\sum m_{\nu}$, to lie in the range 0.10 eV $\lesssim \sum m_\nu \lesssim 0.13$ eV. This is well within reach of the Simon Observatory, which projects to measure the neutrino mass sum with a sensitivity of 40 meV~\cite{SimonsObservatory:2019qwx}. The effective Majorana mass, $m_{ee}$, is predicted to be in a range of $20-40$ meV, putting it within reach of the planned phase-II of the LEGEND experiment, which is projected to constrain $m_{ee} \lesssim  13-29$ meV~\cite{LEGEND:2017cdu}. The effective electron neutrino mass, $m_{\nu_e}^{eff}$, on the other hand, lies between $48-53$ meV,  which is an order of magnitude below the projection $m_{\nu_e}^{eff} < 200$ meV at 90\% C.L. of both the KATRIN and the HOLMES experiments~\cite{Aker:2019uuj,Alpert:2014lfa}. 

The family dependent $Z_2$ symmetry in the lepton sector give rise to flavor violating couplings for the neutral scalars $H$ and $A$. In the decoupling limit, the LFV couplings of $H$ and $A$ are given by $\hat Y$ described in Sec.~\ref{sec:model}. In particular, the LFV coupling $\hat{Y}_{e\mu}$ is enhanced in the low $\tbeta$ part of the viable parameter space identified by neutrino data. Hence, provided that $H$ and $A$ are a few TeV or lighter, $\mu\to e\gamma$ and $\mu\to e$ conversion measurements firmly rule out the low $\tbeta$ region. For the high $\tbeta$ region, next generation experiments are capable of probing it. The Mu2e experiment is expected to reach a sensitivity of $3\times10^{-17}$ on the $\mu\to e$ conversion rate~\cite{Bernstein:2013hba}, which is sensitive enough to probe part of our parameter space, see Fig.~\ref{fig:lowenergylfv}. On the other hand, the MEGII experiment is projected to reach $6\times10^{-14}$ sensitivity on the branching ratio $Br(\mu\to e\gamma)$~\cite{MEGII:2018kmf}, just above the expected branching ratio in our scenario. 

In the case where only $\Phi_2$ couples to quarks, especially when $H$ and $A$ are light enough, e.g., $m_{H/A}\lesssim200$ GeV, LHC searches on $H/A\to \mu^+\mu^-$ provide a more stringent constraint on the parameter space of interested than those of LFV, thanks to the enhanced $\hat{Y}_{\mu\mu}$ coupling, see Fig.~\ref{fig:yhat}. By recasting the CMS search for a Higgs decaying to $\mu^+\mu^-$ in supersymmetric context, we found that LHC data typically exclude  part of the relevant parameter space, see Fig.~\ref{fig:higgstomumu}, which corresponds to region having somewhat lower $\tbeta$ (i.e., red points, see Fig.~\ref{fig:viableparam}).  It is interesting to note that the CMS search has been performed on a small subset of Run II data. In the event that CMS analyzes the full dataset, they could rule out a light Higgs in this particular scenario, or even discover it. 

Finally, we note that the neutrino mass scale is related to the charged scalar mass. Parametrically, $m_\nu$ scales as $\mu \hat f m_\ell^2/m_{H^+}^2$, where $m_\ell$ and $m_{H^+}$ are the charged lepton and charged scalar masses. The tri-linear coupling $\mu$ and the Yukawa coupling $\hat f$ cannot be arbitrary large since the former leads to fine-tuning on the light Higgs boson mass, while the latter is constrained by perturbativity. Specifically, taking $m_\nu\sim10$ meV, $\mu\le1.5$ TeV~\cite{Cai:2017jrq} and $|\hat f|<4\pi$, we obtain $m_{H^+}\lesssim100$ TeV.


\appendix
\section{Diagonalization of the charged lepton mass matrix}
\label{app:diag}
Here we present the expression for parameters of mixing matrices $O_{L,R}$ that diagonalize the charged-lepton mass matrix, as in Eq.~\eqref{mleptond}. Those parameters are defined according to the decomposition given in Eq.~\eqref{eq:ol}, with $\theta_i \to \alpha_i$ for $O_R$. As mentioned in the main text, four of the mixing parameters can be expressed in terms of the other two, chosen to be $\theta_1$ and $\alpha_3$. Their expressions are given below
\begin{align}
s_{\theta_3} &= \frac{s_{\alpha_3}m_\mu m_\tau - c_{\alpha_3}s_{\alpha_2}s_{\theta_1}c_{\theta_1}(m_\tau^2-m_\mu^2)}{\sqrt{c_{\alpha_3}^2(c_{\theta_1}^2m_\tau^2+s_{\theta_1}^2m_\mu^2)\left(s_{\alpha_2}^2(c_{\theta_1}^2m_\tau^2 + s_{\theta_1}^2m_\mu^2) + c_{\alpha_2}^2m_e^2\right)+\left(s_{\alpha_3}m_\mu m_\tau - c_{\alpha_3}s_{\alpha_2}s_{\theta_1}c_{\theta_1}(m_\tau^2-m_\mu^2)\right)^2}},\\
s_{\theta_2} &= \frac{m_\tau c_{\theta_1}(c_{\alpha_3}s_{\alpha_2}c_{\alpha_1}-s_{\alpha_3}s_{\alpha_1}) + m_\mu s_{\theta_1}(c_{\alpha_3}s_{\alpha_2} s_{\alpha_1}+s_{\alpha_3}c_{\alpha_1})}{\sqrt{\left(m_\tau c_{\theta_1}(c_{\alpha_3}s_{\alpha_2}c_{\alpha_1}-s_{\alpha_3}s_{\alpha_1}) + m_\mu s_{\theta_1}(c_{\alpha_3}s_{\alpha_2} s_{\alpha_1}+s_{\alpha_3}c_{\alpha_1})\right)^2+c_{\alpha_3}^2c_{\alpha_2}^2m_e^2}},\\
s_{\alpha_1} &= \frac{s_{\theta_1}m_\mu}{\sqrt{s_{\theta_1}^2m_\mu^2 + c_{\theta_1}^2m_\tau^2}}, \\
s_{\alpha_2} &= -\frac{c_{\alpha_3}s_{\theta_1}c_{\theta_1}m_e^2(m_\tau^2-m_\mu^2)}{s_{\alpha_3}m_\mu m_\tau(c_{\theta_1}^2m_\tau^2+s_{\theta_1}^2m_\mu^2-m_e^2)}.
\end{align}

\begin{acknowledgments}
The work of RP was supported by the Parahyangan Catholic University under grant no. 
III/LPPM/2022-02/71-P.
The work of JJ was supported in part by the Indonesian Institute of Sciences (which later became the National Research and Innovation Agency) through Research Support Facility Program. 
PU would like to thank P. Ongmongkolkul for helpful discussions on the iminuit package~\cite{iminuit}.  PU also acknowledges the National Science and Technology Development Agency, National e-Science Infrastructure Consortium, Chulalongkorn University, and the Chulalongkorn Academic Advancement into Its 2nd Century Project (Thailand) for providing computing infrastructure that has contributed to the results reported within this paper. The work of PU was supported in part by the Srinakharinwirot University under grant no.~035/2565 and the Thailand Toray Science Foundation. 

\end{acknowledgments}

\bibliography{references} 

\end{document}